\documentclass[prl,onecolumn,superscriptaddress,nofootinbib]{revtex4-1}
\usepackage{graphicx}
\usepackage{color}
\usepackage{amsmath}
\usepackage{amsfonts}
\usepackage{mathrsfs}

\begin{document}
Submitted to \textit{Biophysical Reviews and Letters}

\title{Conjecture on the lateral growth of type I collagen fibrils}

\author{Jean Charvolin}
\affiliation{Laboratoire de Physique des Solides (CNRS-UMR 8502), B{\^a}t. 510, Universit{\'e} Paris-sud, F 91405 Orsay cedex}

\author{Jean-Fran\c cois Sadoc}
\email{sadoc@lps.u-psud.fr}
\affiliation{Laboratoire de Physique des Solides (CNRS-UMR 8502), B{\^a}t. 510, Universit{\'e} Paris-sud, F 91405 Orsay cedex}

\begin{abstract}
Type I collagen fibrils have circular cross sections with radii mostly distributed in between $50$ and $100$ nm and are characterized by an axial banding pattern with a period of $67$ nm. The constituent long molecules of those fibrils, the so-called triple helices, are densely packed but their nature is such that their assembly must conciliate two conflicting requirements : a double-twist around the axis of the fibril induced by their chirality and a periodic layered organization, corresponding to the axial banding, built by specific lateral interactions. We examine here how such a conflict could contribute to the control of the radius of a fibril. We develop our analysis with the help of two geometrical archetypes : the Hopf fibration and the algorithm of phyllotaxis. The first one provides an ideal template for a twisted bundle of fibres and the second ensures the best homogeneity and local isotropy possible for a twisted dense packing with circular symmetry. This approach shows that, as the radius of a fibril with constant double-twist increases, the periodic layered organization can not be preserved without moving from planar to helicoidal configurations. Such changes of configurations are indeed made possible by the edge dislocations naturally present in the phyllotactic pattern where their distribution is such that the lateral growth of a fibril should stay limited in the observed range. Because of our limited knowledge about the elastic constants involved, this purely geometrical development stays at a quite conjectural level.
\end{abstract}

\maketitle

%%%%%%%%%%%%%%%%%%%%%%%%%%%%%%%%%%%%%%%%%%%%%%%%%%%%%%%%%%%%%%%%%%%

\section{Introduction }

Type I collagen fibrils, cable-like assemblies of long biological molecules, the so-called triple helices, are the major constituents of connective tissues. Electron microscopy and atomic force microscopy provide images of fibrils which show that :\\
-	they are not smooth, as most of the bundles of fibres built by biopolymers, but are striated all along their length with a constant period $s=67$ nm,\\
-	 the radii of their circular cross sections are in general distributed in a range going from $50$ to $100$ nm, with a few exceptions of about $150$ or $230$  nm.

These characteristics are common to all type I collagen fibrils whatever their origin, conditions of extraction, preparation and observation, in vitro as well as in vivo. This suggests that the lateral size of a fibril, once its growth has been triggered by external biological factors, is mostly controlled by the evolution of its internal structure. In other words, the free energy of a fibril as function of its radius $\rho$ should contain, in addition to the sum of the cohesive and interfacial
terms $-a \rho^2 + b\rho$, whose minima correspond to molecular dispersion or mass precipitation only,
a third intrinsic term $+c \rho^n$ with $n>2$ to avoid mass precipitation and limit the growth.

\begin{figure}[tpb]
%$$\resizebox{1\columnwidth}{!}{%
%
\includegraphics{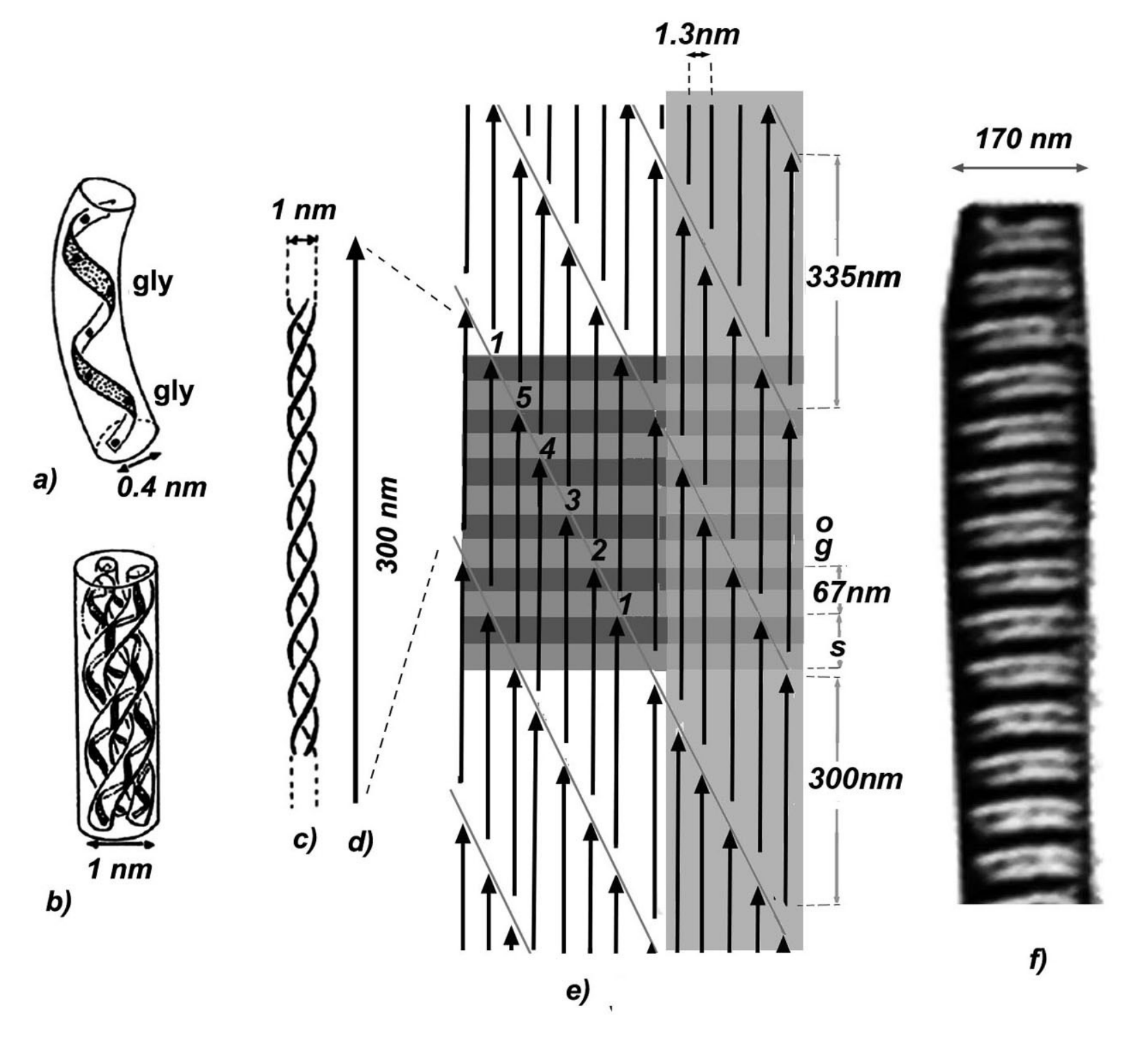}
%}$$
%
\caption{Left-handed collagen molecules (a) assemble in right-handed triple helices with rotation symmetry of order 3 (b,c) represented by long arrows (d) which, in turn, are proposed to be assembled with a regular shift (e), the so-called Hodge-Petruska staggering, in order to create regions of gap ``g'' and overlap ``o'' giving account of the axial striations of the fibril (f).}

\label{f1}
\end{figure}

The first candidate to be thought of is of course the double-twist induced by the chirality of the triple helices as its propagation in our Euclidean space generates elastic stresses which can play such a role. This has been shown   for smooth bundles of fibres ordered along a hexagonal lattice \cite{grason}, but collagen fibrils are striated and their triple helices, although densely packed, do not present such an order \cite{charvolinsadocBRL}.
The striations shown by the fibrils result from  specific interactions between triple helices leading to a regular shift, the Hodge Petruska (HP) staggering, with the alternate  overlap and gap regions shown in figure~\ref{f1}. The segment of triple helices are densely organized in the overlap regions, but not in the gap regions where one segment out of five is replaced by a vacancy. The distribution of those vacancies in the plane of figure~\ref{f1} and out of it determines a layered periodic structure made visible by the striations.
%
%______________________ Fig.1___________________________________
%
%________________________________________________________________
%
%
Such a periodic layering is not compatible with a double-twist which would impose a variation of the layer thickness. This problem was addressed in the case of smectic phases of chiral mesogenic molecules and the solution proposed was that of a double twist texture with the local smectic order preserved in coaxial domains separated by cylindrical screw dislocation walls \cite{kamien}.  However, the layers of type I collagen fibrils are different from those of liquid crystalline smectic phases. Owing to the fact that they are built by triple helices much longer than the layer thickness, these layers are not liquid, can not glide on each other and their number along a fibril is imposed by that on its axis.

We examine here how a system of long triple helices could conciliate double-twist and layering in the best manner possible analyzing the geometrical distortions imposed by the coexistence of these two terms. In order to consider only the perturbation brought by the layering, without interfering with those related to the propagation of the double-twist in our Euclidean space $R_3$ and molecular extension, we first build a template using the Hopf fibration of the hypersphere $S_3$  whose fibres of constant length are organized with a uniform double-twist (see appendix A). We also describe the transverse dense organization of the triple helices in this template with the algorithm of phyllotaxis which ensures the best packing efficiency in a situation of circular symmetry (see appendix B). We finally show that the interplay between the development of an axial layering in presence of a constant double-twist and the edge dislocations naturally present in a phyllotactic pattern can limit the growth of the template built in $S_3$ and that this is maintained when this curved space is projected onto the flat space $R_3$.

\section{ Geometrical  template built with the Hopf fibration in $S_3$}

\subsection{Layered structuring}

\begin{figure}[tpb]
%$$\resizebox{1\columnwidth}{!}{%
%
\includegraphics{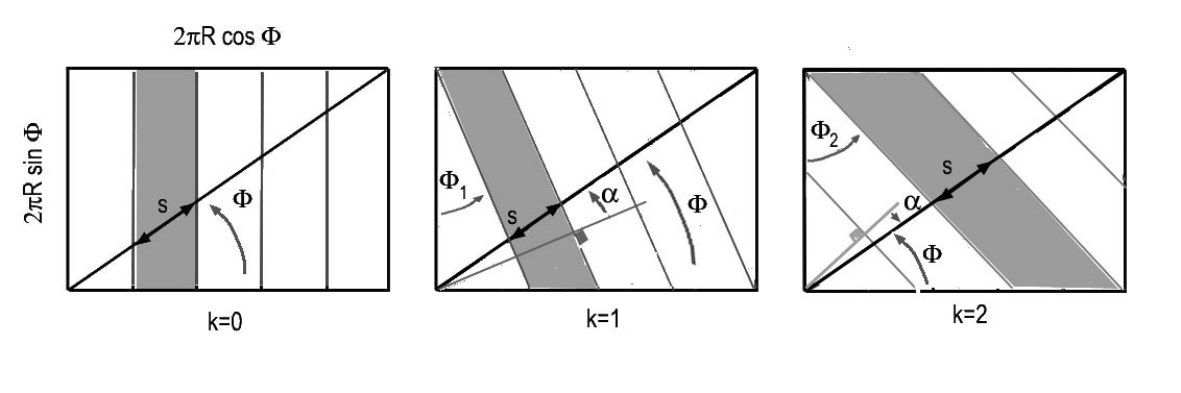}
%}$$
%
\caption{ Rectangle representative of torus $\phi$ supporting fibres at an angle $\phi$ from the torus axis $\phi =0$ with the traces of layers normal to this axis (a) and those of helicoidal layers with pitches of one (b) or two (c) strips, the tilt angle $\alpha=\phi-\phi_k$ and $2\pi R/s=n=5$ on this figure.}

\label{f2}
\end{figure}

The fibres of the Hopf fibration in a hypersphere $S_3$ of radius $R=P/2\pi$, where $P$ is the pitch of the double twist, are great circles of length $2 \pi R$. Several periodic configurations of layers can be superposed on this fibration, either the simple stacking of flat layers normal to the $C_\infty $ axis $ \phi=0$ of the fibration or helicoidal layers tilted relative to this axis \footnote{Helicoidal layers in $S_3$ are indeed determined by surfaces whose distance varies when $ \phi$  increases, but this occurs largely beyond the values considered here.}. The traces of such layers determine strips on the rectangle representative of torus $\phi$  supporting the fibres at an angle $\phi$  from the axis, as drawn on figure~\ref{f2}.  The strips must be drawn in order to respect the continuity of the layers when identifying two by two the sides of the rectangle to build the torus. Moreover, being connected by the triple helices crossing them, their number $n$ of intersections with each fibre, or diagonal, must stay constant equal to that $n=2 \pi R/s$ measured along the $ \phi=0$ axis with layers normal to it.

%
%______________________ Fig.2___________________________________
%
%________________________________________________________________
%
%

\begin{figure}[tpb]
%
%$$\resizebox{1\columnwidth}{!}{%
\includegraphics{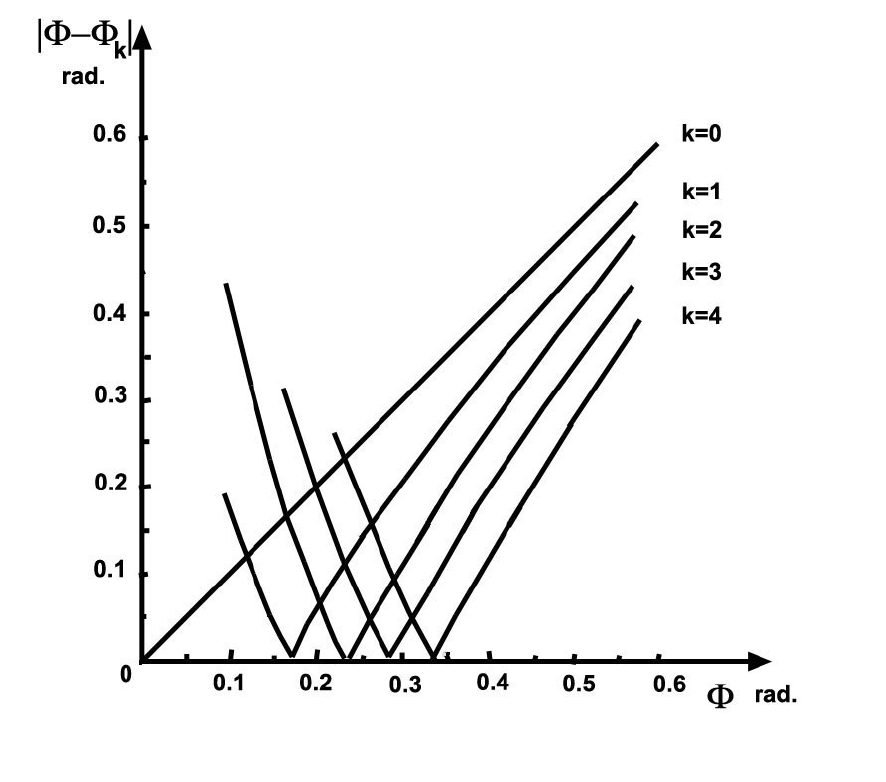}
%}$$
%
\caption{Variations of the tilt $|\phi-\phi_k|$ of the fibres with respect to the layer normal as $\phi$ increases in configurations $k$ for $n=2 \pi R/s=36$ corresponding to a double-twist pitch $P=2 \pi R$ of about $2400$ nm.}

\label{f3}
\end{figure}

These periodic configurations are labeled according to the number $k=0,1,2,\ldots$ of strips defining the pitch of the helix drawn on torus $\phi$. If the number $n$ of strips intersecting the diagonal is the same for all of them, their numbers of intersections with the horizontal and vertical sides of the rectangle vary as $n-k$ and $k$ respectively. The tilt of the fibres relative to the layer normal can then be written as $|\phi-\phi_k|$  where  $\phi_k=\arctan( k/ ((n-k)\tan \phi ))$ and its variations are shown in figure~\ref{f3} for a double-twist pitch $P=2400$ nm deduced from the observations described in a next section.

%%
%
%______________________ Fig.3___________________________________

\begin{figure}[tpb]
%
%$$\resizebox{1\columnwidth}{!}{%
\includegraphics{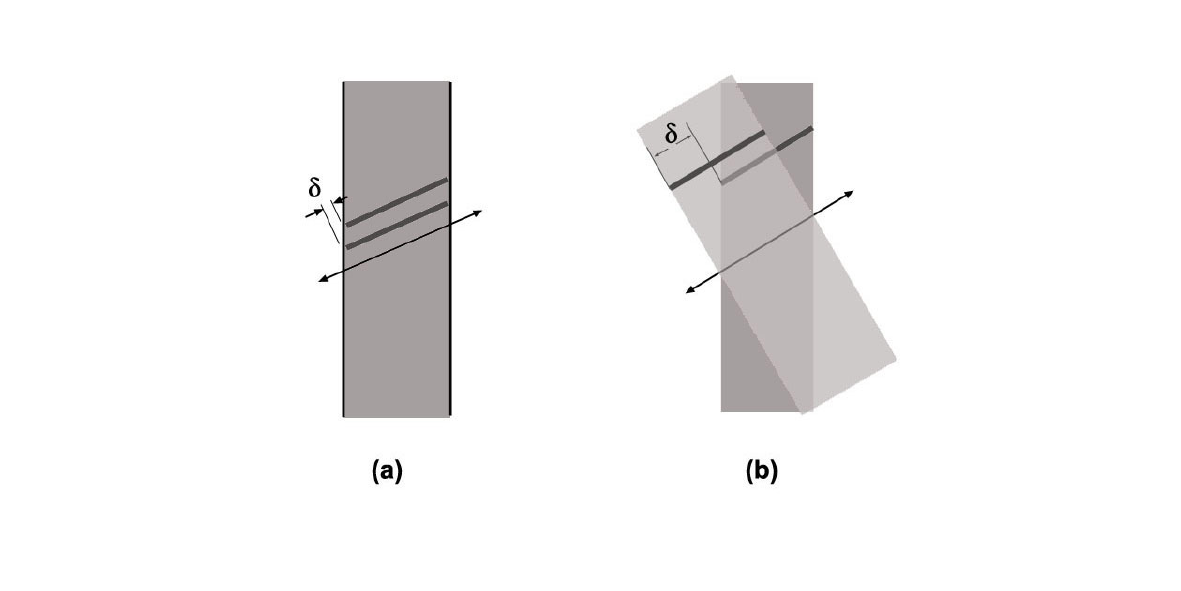}
%}$$
%
\caption{Relative displacements $\delta$ or shears required when the triple helices are tilted with respect to the layer normal (a) or when a change of configuration occurs (b), the black lines represent the direction of the triple helices.}
\label{f4}

\end{figure}

For $k=0$ the tilt of the fibres increases linearly, for $k>0$ it decreases to zero for $ \phi=\arctan \sqrt{k/(n-k)}  $. Those curves show that too large an increase of the tilt as $\phi$ increases in one configuration $k$ can be avoided changing for configuration $k+1$.
These variations in one configuration, or when moving from one configuration to the following, should be favored or not by the local organization of the triple helices as they imply a shear of the HP staggering along their common direction as shown in figure~\ref{f4}.

\subsection{Constraints introduced by the HP staggering}
The proposition of the HP staggering was advanced considering that lateral chemical bonds between triple helices are able to build the regular shift  shown on figure~\ref{f1}. Owing to the fact that triple helices have a rotation symmetry of order 3, similar drawings can be built in planes at $\pi/3$ and $2\pi/3$ from that of figure~\ref{f1} so that the gap and overlap regions be distributed coherently along the fibril axis to build the layering at the origin of the striations\footnote{As drawn in figure~\ref{f1}, the HP staggering would develop bidimensional sheets of large extension, it was indeed first proposed that each period of the HP staggering closes onto itself forming a cylindrical microfibril containing five triple helices and that such microfibrils assemble to build a fibril, this process is now questioned.}. The interactions stabilizing this organization certainly restrain the relative displacements of triple helices along their lengths required if the layered structuring in the Hopf fibration evolves as described above. The energy cost associated with shears could be lowered, eventually suppressed, if the propagation of lateral bonds between triple helices was interrupted so that they can slide along each other. The phyllotactic model recently proposed to describe the dense packing of triple helices in fibrils with circular cross section \cite{charvolinsadocBRL} provides such an opportunity.

%
%______________________ Fig.4___________________________________
%
%________________________________________________________________
%
%

\subsection{Phyllotaxis and growth control in $S_3$}
\begin{figure}[tpb]
%$$\resizebox{1\columnwidth}{!}{%
%
\includegraphics{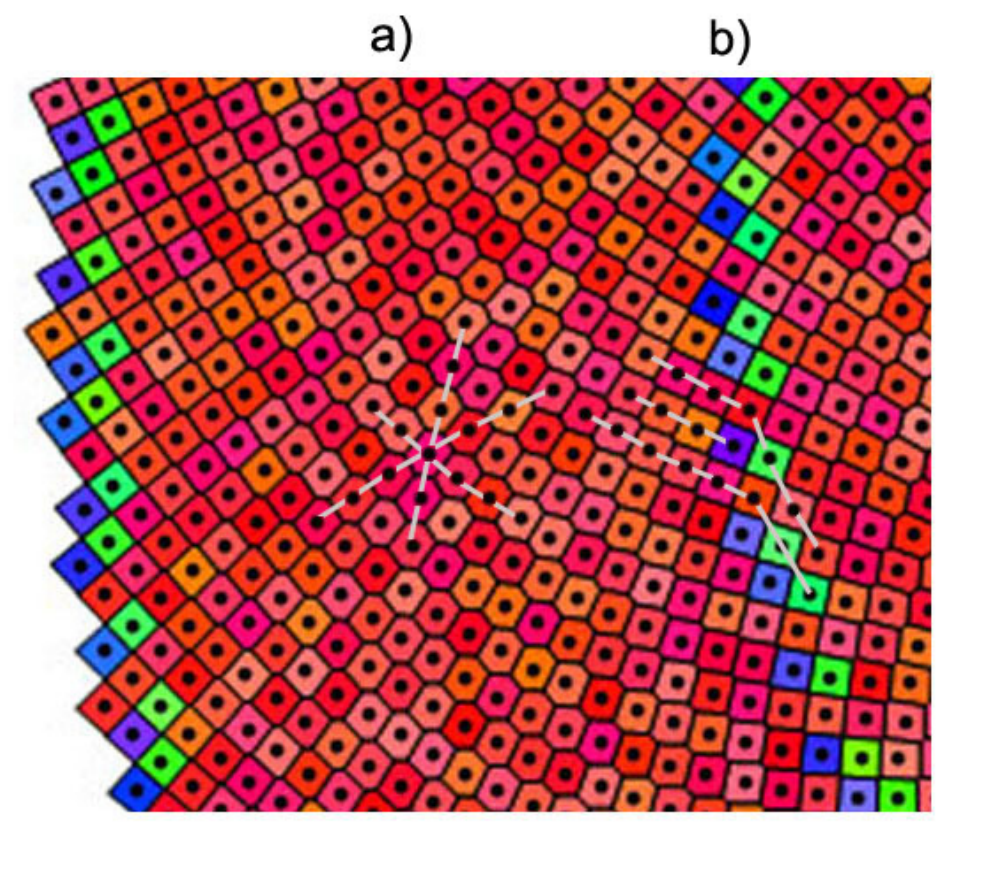}
%}$$
%
\caption{The Voronoi cells are hexagonal in the core of a grain (a) but are distorted with a shape quite close to that of a square or build dipoles of blue pentagons and green heptagons in a grain boundary (b), those dipoles are dislocations introducing new parastiches on the grain boundary, clear  lines are parastiches.}
\label{f5}
\end{figure}

The distribution of points representing the Hopf fibration on its spherical basis is that of the phyllotactic pattern built on this surface by the algorithm of phyllotaxis. The representative points of closely bonded triple helices in a HP staggering are to be aligned along the parastiches of this pattern, the lines of shortest distances between neighbor points, and so are the interactions building this HP staggering. The cohesion of the assembly should therefore be strong within hexagonal grains where the symmetries of the triple helices and their environment are coherent, each point being at the intersection of three parastiches in a hexagonal Voronoi cell with six close neighbors. In the vicinity of grain boundaries, where the hexagons are strongly distorted towards a shape quite close to that of a square or are transformed into heptagons or pentagons, the molecular and local symmetries are no longer coherent.  The parastiches, and the propagation of the interactions aligned along them are perturbed along grain boundaries, as shown in figure~\ref{f5}, and the cohesion of the assembly should be weakened.

%
%
%______________________ Fig.5__________________________________
%
%________________________________________________________________
%
%

\begin{figure}[tpb]
%$$\resizebox{1\columnwidth}{!}{%
%
\includegraphics{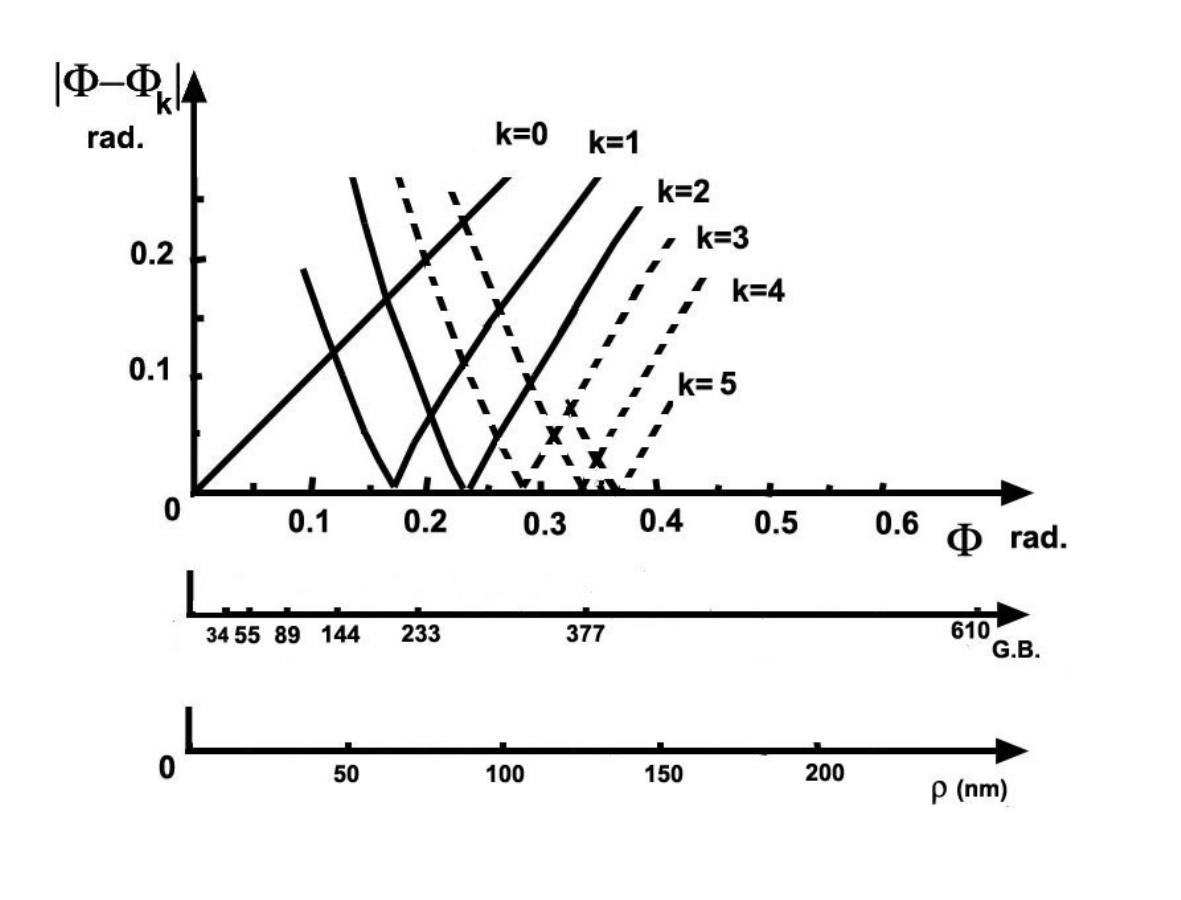}
%}$$
%
\caption{Positions of the cancelation points of the tilt as function of  $\phi$ compared with that of the grain boundaries (GB) and with the radius $\rho =R \phi$ of the torus supporting the fibres. The positions of the grain boundaries, identified by their number of dislocations, are obtained from a phyllotactic pattern built on the spherical basis of a Hopf fibration with the double-twist pitch $P=2400$ nm and a distance between site $d=1.3$ nm.}
\label{f6}
\end{figure}

Triple helices would therefore be free to slide along each other, making changes of configurations possible, on torii whose generator circles are grain boundaries. As those changes are also expected to take place in between two cancelation points of the tilt, the positions of the grain boundaries are compared with those of the cancelation points in figure~\ref{f6}.

The core of the pattern, up to grain boundary 21, shows a high density of defects and therefore has a low cohesion. A simple tilt can grow easily without need for a change of configuration. When $\phi$ increases beyond this core, the defects concentrate in well individualized grain boundaries delimiting grains without defects, hence with a higher cohesion. This constrains the growth of the tilt and calls for a change of configuration on the first grain boundary met. However, while the separation between two consecutive cancelation points of the tilt decreases that of grain boundaries increases and changes of configuration become less and less possible as $\phi$ increases.

%
%
%______________________ Fig.6__________________________________
%
%________________________________________________________________
%
%
%
For instance, grain boundaries $144$ and $233$ would favor the changes of configuration $k=0$ to $k=1$ and $k=1$ to $k=2$ respectively, but grain boundary $377$ is too far to favor that from $k=2$ to $k=3$ and even that from $k=3$ to $k=4$. In that zone, where configuration changes beyond $k=2$ at the most are hindered, the tilt should be forced to increase almost linearly with $\phi$, so that the energy associated with it, $2\pi R^2\int\phi^2\phi d\phi $, would vary as $\phi^4$, or $\rho^4$. This would provide the term which, being added to those of volume and surface varying as $-\rho^{-2}$ and $\rho$ mentioned  in the introduction, is needed to limit the growth of the fibrils. From figure~\ref{f6}, this limitation would occur around a radius $\rho$ of about $80$ nm for the template built in a hypersphere $S_3$ containing a Hopf fibration with a double-twist pitch $P=2400$ nm. A value of $\rho$ which corresponds to most of the radii observed in our space $R_3$.

\section{ From the virtual template in $S_3$ to real materials in $R_3$ }

The stereographic projection is the simplest way to carry out the transfer from a curved space to a flat one. This conformal projection does not affect the topology, a torus in $S_3$ is projected as a torus in $R_3$. Choosing the pole of projection on the  $\phi=\pi/2$ axis and a projection space containing the  $\phi=0$ axis in one coordinate plane, as in figure~\ref{f9} of appendix A, this axis stays unchanged and the radius of the generator circle of the projection behaves as $R \tan \phi $  which stays close to $\rho =R \phi$ in $S_3$  for the small values of $\phi$ considered here  \cite{charvolinsadocLivre}. This enables a direct comparison with the fibrils described below.

%
%______________________ Fig.7__________________________________
%
\begin{figure}[tpb]
%$$\resizebox{1\columnwidth}{!}{%
%
\includegraphics{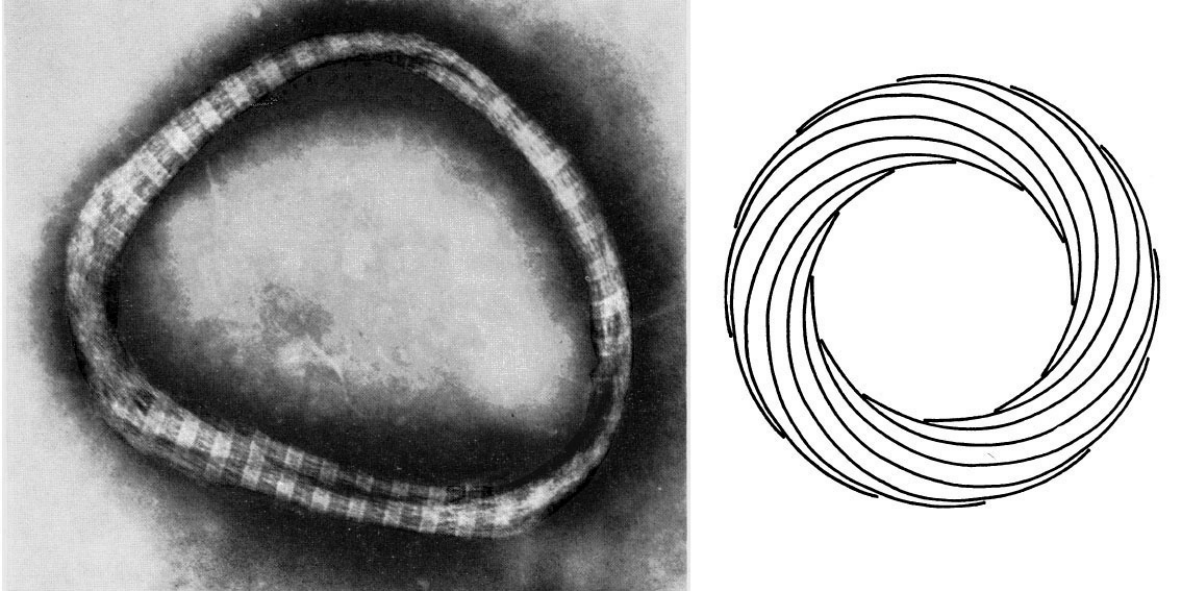}
%}$$
%
\caption{Electron micrograph of a toroidal fibril and its interpretation, from  \cite{cooper}.}
\label{f7}
\end{figure}
%________________________________________________________________
%
%

\subsection{Toroidal fibrils}
This remarkable morphology shown in figure~\ref{f7} was obtained by precipitation of calf-skin tropocollagen in vitro  and studied by electron microscopy \cite{cooper}.

Striations as well as longitudinal traces making a complete turn around the director circle of the torus are quite discernable on the micrographs. The second shows the existence of a double-twist with the topology of a Hopf fibration. The period of the striations is of $67$ nm, as for any fibril, and the perimeter of the director circle, which is therefore the pitch of the double-twist, varies from $P=2300$ to $2600$ nm, hence our choices of $400$ nm for the radius $R=P/2\pi$ of $S_3$ and the number $n=36$ of layers created by the HP staggering in the preceding sections. The radii of the cross sections of those toroidal fibrils, when they can not be suspected to have been disrupted along their perimeter during drying, lies in between $50$ and $100$ nm in good agreement with the above expectation.

Toroidal fibrils can also be formed by other biopolymers such as polypeptides having diameters close to that of collagen triple helices  \cite{blaisgeil}. An organization of fibres having the topology of the Hopf fibration is clearly visible on some micrographs shown in this reference, but always without any striation, hence without layering. Comparing toroidal fibrils with equivalent director circles, the radii of the generator circles of these built by polypeptides are more than two times larger than that those built by collagen, as if, owing to the absence of layering, the growth of the first was not limited so early than that of the second.

\subsection{Straight  fibrils}
In the absence of a simple method to transform a torus in $S_3$ into an infinite cylinder in $R_3$, we just transfer the organization of fibres and layers built in the toroidal template along a straight cylinder whose director circle has the radius of the generator circle of the torus. As straight fibrils in tendons have been the objects of many X-ray scattering structural studies, this open the possibility to confront the results of the model with those observations.

\begin{figure}[tpb]
%$$\resizebox{1\columnwidth}{!}{%
%
\includegraphics{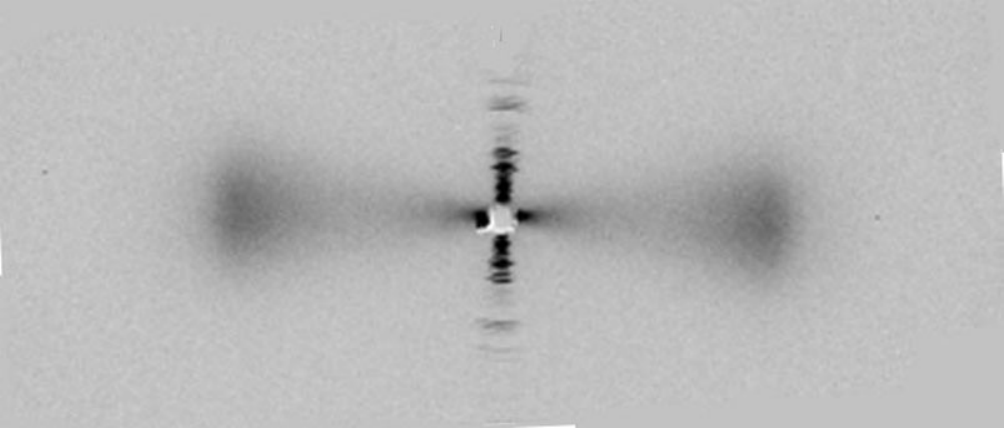}
%}$$
%
\caption{X-ray microdiffraction pattern of a single mouse tail tendon, from  \cite{doucetbriki}. }
\label{f8}
\end{figure}

First, the lateral size obtained along this approach is such that the fibril cross section is mostly occupied by the domain of the phyllotactic pattern proposed in  \cite{charvolinsadocBRL}  in which the long range ordering of triple helices is compatible with that deduced from X-rays scattering studies. Second, X-ray diffraction patterns are also characterized by scatterings with fan-like shapes along the equatorial and meridian directions  \cite{wesshammersley,mcbride,wessorgel,laingorgel,doucetbriki}. An example is shown in figure~\ref{f8}.

%
%
%______________________ Fig.8__________________________________
%
%________________________________________________________________
%
%
%
In most of the studies, the fans around the equator, to be associated with the transverse organization of triple helices, are rather diffuse with opening angles from $15 ^\circ$ to $25 ^\circ $. and those around the meridian, to be associated with the longitudinal striations, is better defined with smaller opening angles from $0 ^\circ$ to $15^\circ$. The opening angles being different, these fan-like shapes can not have their origins in the disorientation of the samples. This suggests that the fibres and the layers are oriented differently, as proposed by our model where the angles $\phi$ and $\phi_k$ may differ. For instance, on the periphery of a fibril with a radius $\rho\approx80$ nm, the triple helices should be oriented at $\phi\approx0.2$ rad or $11^\circ$ from the fibril axis whereas the layers of configuration $k=1$ should be oriented at $\phi_1\approx0.14$~rad or $8^\circ$. Also, while the orientation
$\phi$ of the fibres increases continuously with the distance from the axis, that $\phi_k$ of the layers vary by jump each time the configuration changes. However, as the tendons contain fibrils of different sizes, a precise relation between angles and radii can not be deduced from these X-rays studies.

\section{Conclusion, beyond the conjecture?}

The molecular interactions determining the assembly of triple helices in type I collagen fibrils are  of rather complex nature and exert their actions in constrained situations. For instance the circular symmetry imposed by the interfacial tension prevents the propagation of a transverse crystalline order and the double-twist configuration associated with the molecular chirality of the triple helices is not compatible with the periodic axial layering issued from their HP staggering. The structure of the fibrils is most likely the result of quite subtle competitions between different terms whose exact knowledge is rather poor at the moment. This situation precludes any quantitative search for compromises and we limited our examination of this structural problem to its geometrical foundation.

We introduced a periodic axial layering into a template in which triple helices of constant length develop a uniform  double-twist and are densely packed according to the spirals of a phyllotactic pattern. As the radius of template increases, the periodic layering can not be preserved without moving from planar to helicoidal configurations around the axis of the template and this requires shearing the triple helices along their common direction. Such relative displacements cannot be obtained without going against the lateral bindings of the triple helices along the shearing surface. Fortunately, edge dislocations naturally present in the spirals of a phyllotactic pattern, along which triple helices are not connected to some of their neighbors, provide such opportunities. Those dislocations are concentrated along circular grain boundaries which are more and more distant as the radius of the template increases so that the periodic layering can not be preserved for a lateral size larger than that shown by real fibrils. Such an intrinsic stress, as well as that generated by the propagation of the double twist  \cite{grason} , would therefore contribute to the control of the growth without calling for external factors of control.

In spite of this agreement, grey areas remain which call for appropriate experimental studies. They concern the internal structure of fibrils with well characterized radii, including the propagation of the HP staggering, and the relation between twist and radius imposed by the template. In addition to the experimental studies already quoted in the article, a few others might open such directions:\\
-	electron microscopy of tendon normal sections showing a spiral lateral    organization of the triple helices  \cite{hulmesjesior},\\
-	atomic force microscopy of fibrils deposited on a substrate showing tilted striations and twisted triple helices at the periphery of fibrils with large radii   \cite{wenger},\\
-	electron microscopy putting in light a growth from pointed paraboloidal tips eventually susceptible to be related to the organization of the HP staggering  \cite{prockopfertala},\\
-	mechanical studies such as those described in  \cite{bozecvdh}, or on individual fibrils with optical twitters, which would give access to the elastic constants needed to develop a thermodynamical approach.

Beyond its eventual interest for the conception of artificial tissues, such a program is also justified as a contribution to the analysis of the respective roles of genetics and physical chemistry in building the morphologies needed for biological materials to fill their functions.

\section{Appendix A: Hopf fibration in S3 }
\begin{figure}[tpb]
%$$\resizebox{1\columnwidth}{!}{%
%
\includegraphics{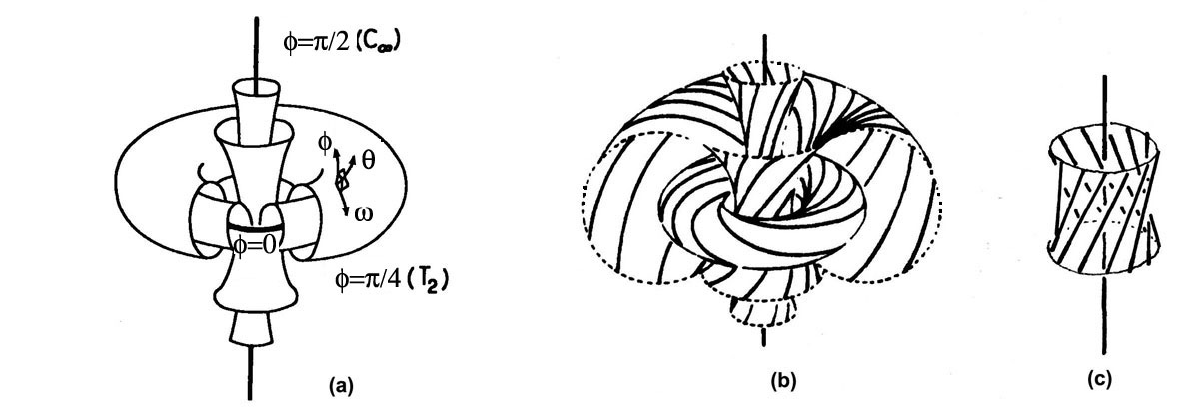}
%}$$
%
\caption{Stereographic projections of the hypersphere $S_3$ with toroidal coordinates (a), of the family of Hopf fibres supported by those torii (b) and the double-twist of the local surrounding along every fibre of the family (c).}
\label{f9}
\end{figure}

\begin{figure}[tpb]
\begin{center}
\includegraphics{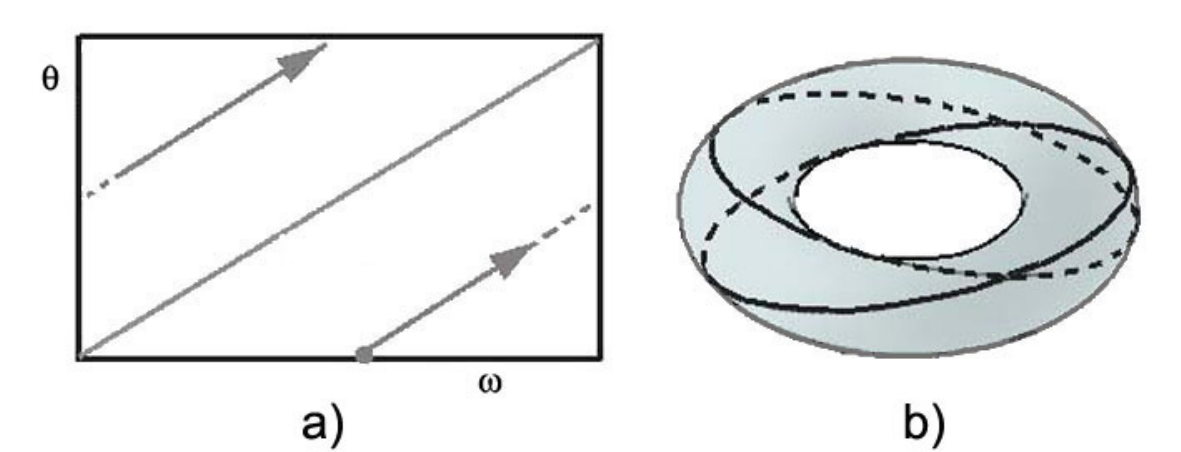}

\caption{A rectangle with edges of lengths $2\pi R cos \phi $ and $2 \pi R sin \phi$ (a) is folded into a torus (b) in $S_3$. A diagonal of the rectangle and one of its parallel lines become two enlaced great circles of length $2\pi R$ on the torus, fibres at an angle   from its  $\phi=0$ axis. }
\label{f10}
\end{center}
\end{figure}

A stereographic projection of the hypersphere $S_3$ with one family of Hopf fibres onto the Euclidean space $R_3$  is shown in figure~\ref{f9}  \cite{sadoccharvolinadn}.
 %______________________ Fig.9__________________________________
%
%________________________________________________________________
%
%
The fibres, great circles of $S_3$, can be drawn on nested parallel torii characterized each by an angle  $\phi$, those with  $\phi=0$ and $\pi/2$ are reduced to great circles and are the $C_\infty$  axes of the fibration. The fibres are enlaced, each one making one turn around the others.  Torus $\phi$  and its fibres are at a distance $R\phi$  from the axis $ \phi=0$ where $R$ is the radius of $S_3$, such a torus can be built in $S_3$ identifying two by two the opposite sides of the rectangle shown in figure~\ref{f10}.
 %______________________ Fig.10__________________________________
%
%________________________________________________________________
%

Following this identification, the fibres of torus  $\phi$, all at the distance $R\phi$  from the axis  $\phi=0$, are all oriented at the same angle  $\phi$ with respect to this axis and make a complete turn around it keeping a constant distance between themselves. If another set of  $\phi=0$ and $\pi/2$ axes is chosen among the great circles of $S_3$ at a distance
$\pi R/2$ from each other, the same situation is reproduced so that any two fibres keep a constant distance between themselves, the fibres are also called Clifford parallels. This is a perfect double-twist configuration with a pitch $P=2\pi R$. The fibres being parallel, their organization in the hypersphere $S_3$ can be simply represented by a distribution of points on a sphere, the basis of the Hopf fibration, in a way similar to the representation of a set of parallel straight lines in $R_3$ by points on a plane $R_2$.

\section{Appendix B: phyllotaxis}

The shape of the cross section of a dense fiber bundle is expected to reflect the symmetry of its molecular packing. However, this statement is belied by type I collagen fibrils which show a circular cross section while structural studies suggest that their molecules can be assembled with some long range lateral order. We recently examined how the iterative process of phyllotaxis, a non conventional crystallographic solution to packing efficiency in situations of high radial symmetry  \cite{sadocRCactacryst}, could establish a link between those two apparently conflicting points \cite{charvolinsadocBRL}.

\begin{figure}[tpb]
%$$\resizebox{1\columnwidth}{!}{%
%
\includegraphics{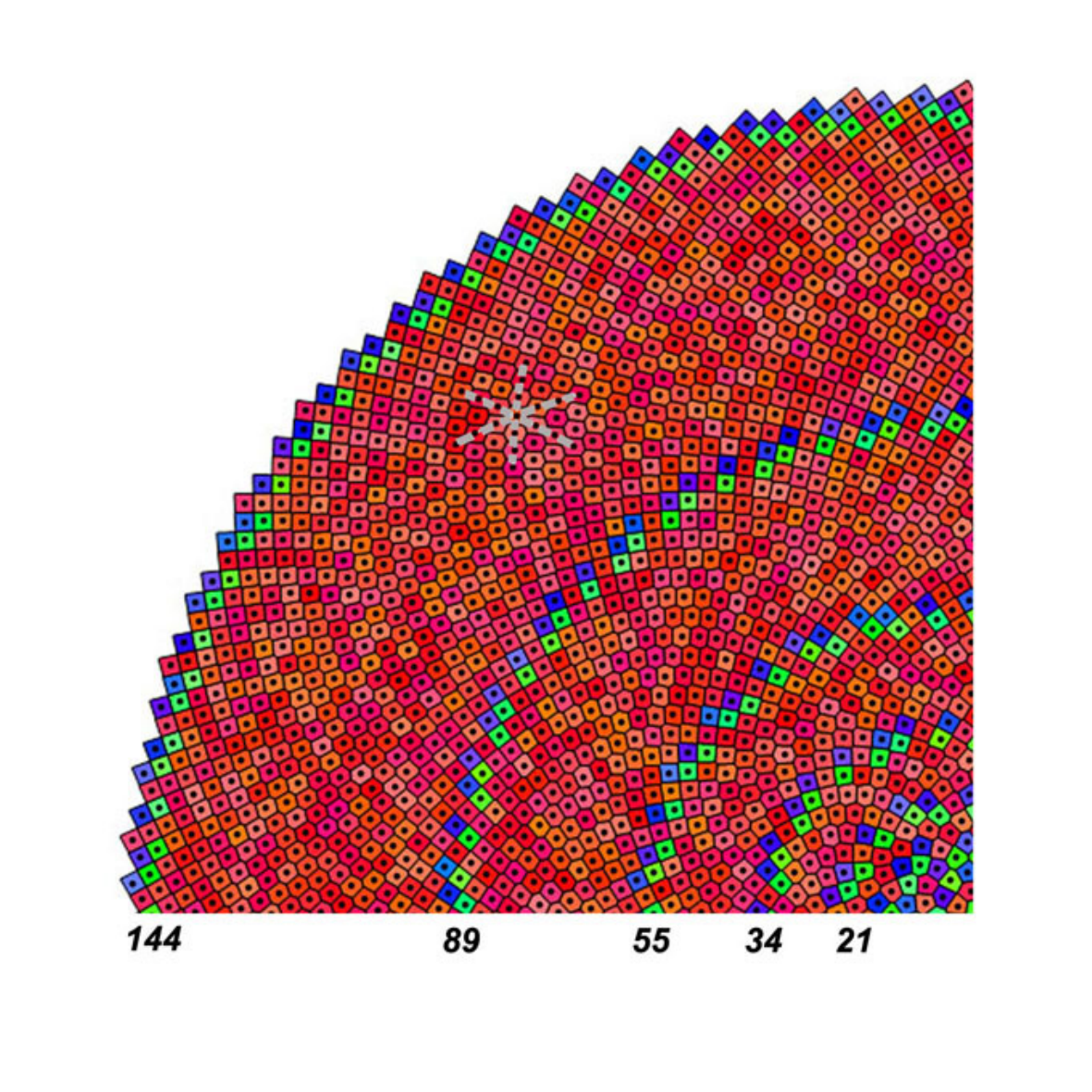}
%}$$
%
\caption{A quadrant of a set of 6500 points organized on a plane according to the algorithm of phyllotaxis with the golden ratio. Each point is surrounded by its polygonal Voronoi cell whose number of sides corresponds to the number of first neighbors around this point. Blue cells are pentagons, orange cells hexagons and green cells heptagons, those polygons are not necessarily regular. Grain boundaries are labeled by Fibonacci numbers corresponding to their number of dipoles. The three spirals joining the first neighbors of the points are called parastichies.}
\label{f11}
\end{figure}

A phyllotactic organization of points indexed by $s$ is built by an algorithm such that the position of point $s$ is given by its polar coordinates $r=a \sqrt{s}$ and  $\theta=2 \pi \lambda s$ that is $r=(a/\sqrt{2 \pi\lambda})\sqrt{\theta}$   which is the equation of a Fermat spiral, the generative spiral. The area of the circle of radius $r$ which contains $s$ points is $\pi a^2 s$ so that the area per point has the value $\pi a^2$, indeed it oscillates close to this value for small s then converges towards it. The most homogeneous and isotropic environment, or the best packing efficiency in radial symmetry, is obtained with   $\lambda=1/\tau$  where  $\tau$ is the irrational golden ratio $(1+ \sqrt{5})/2$  \cite{ridley}. A sector of a phyllotactic pattern for $N=6500$  points on a plane with their Voronoi cells is shown in figure~\ref{f11}.
 %______________________ Fig.11__________________________________
%
%________________________________________________________________
%

\begin{figure}[tpb]
%$$\resizebox{1\columnwidth}{!}{%
%
\includegraphics{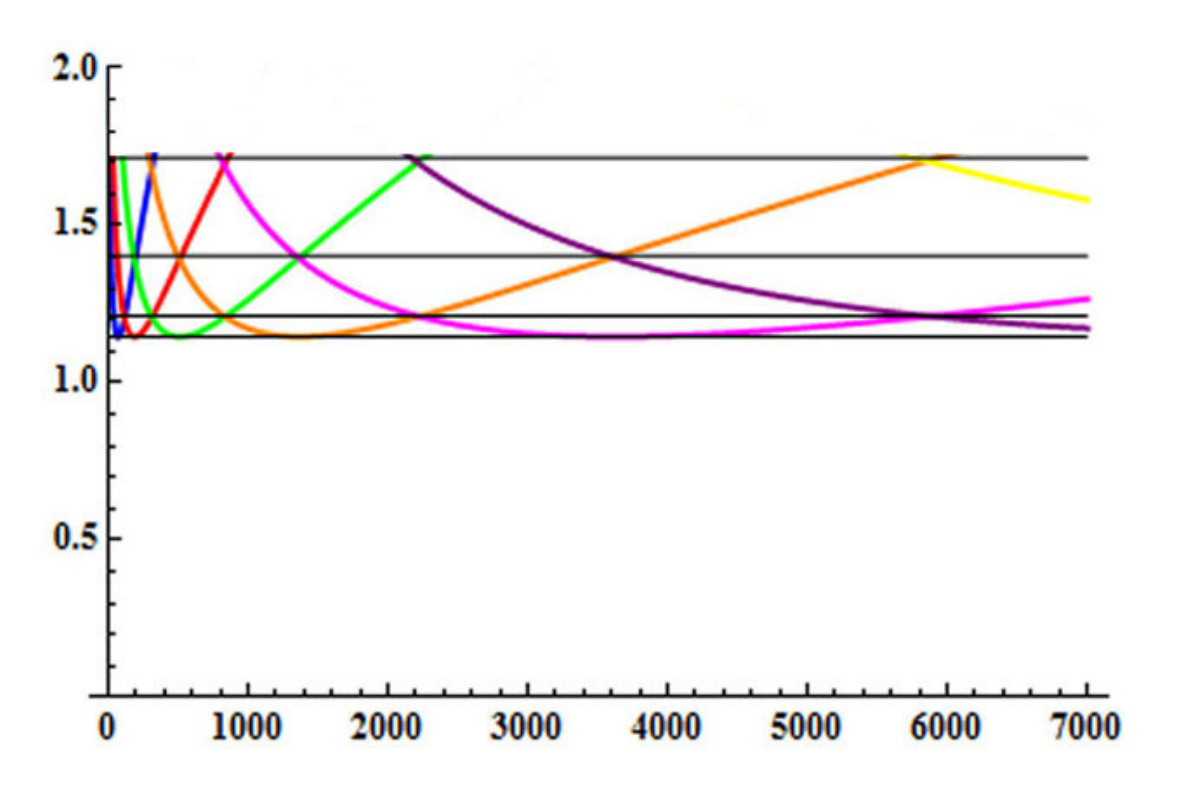}
%}$$
%
\caption{Figure 12. Distances between first neighbor points with the $a$ parameter adjusted to have a mean distance close to $1.3$ nm. The three colours correspond to the three parastichies, the upper and lower crossings on the same verticals correspond to grain boundaries and the intermediate ones to the cores of hexagonal grains.}
\label{f12}
\end{figure}

In such a pattern, pentagons and heptagons are topological defects distributed among the hexagons. They appear concentrated in narrow circular rings with constant width separating large rings of hexagons whose width increases as one moves from the core towards the periphery. In the narrow rings, pentagons and heptagons are associated in dipoles separated by hexagons whose shape is close to that of a square with two corners cut. The rings of dipoles are indeed grain boundaries separating hexagonal grains and the dipoles are dislocations introducing the new parastichies needed to maintain the density as constant as possible. The radii of those rings of dipoles tend to follow the Fibonnacci series, $f_u=f_{u-1}+f_{u-2}$ from $f_0=0$ and $f_1=1$, which makes the organization self-similar, invariant by a change of scale $\tau^n$.
The evolutions of the distances between first neighbor points, as measured along the three parastichies,  are shown in figure~\ref{f12}.

 %______________________ Fig.12__________________________________
%
%________________________________________________________________
%

The algorithm of phyllotaxis has also been developed on the sphere with the aim of applying it for the spherical basis of the Hopf fibration. Patterns of concentric grains separated by circular grain boundaries similar to that described on the plane are obtained and the perimeters of the grain boundaries are independent of the curvature. It results from this last point that the distances of the grain boundaries to the pole measured on the sphere vary with its curvature~\cite{jfsjcnrcurve}.

%\section*{References}

\end{document}